\documentclass[twocolumn,aps,prl,superscriptaddress]{revtex4-2}

\usepackage{amsmath}
\usepackage{graphicx}
\usepackage{MnSymbol}

\begin{document}

\title{On opening crack propagation in viscoelastic solids}

\author{B.N.J. Persson}
\affiliation{PGI-1, FZ J\"ulich, Germany, EU}
\affiliation{www.MultiscaleConsulting.com}

\begin{abstract}
We show that the Persson-Brener theory of crack propagation in viscoelastic
solids gives a viscoelastic fracture energy factor 
$G/G_0 = 1+f$ which is nearly the same as the viscoelastic factor obtained using the 
cohesive-zone model.
We also discuss finite-size effects and comment on the use of crack propagation
theories for  ``solids'' with a viscoelastic modulus that vanishes at zero frequency.
\end{abstract}

\maketitle

\setcounter{page}{1}
\pagenumbering{arabic}




{\bf Introduction}

Crack propagation in viscoelastic solids, or at the interface between a viscoelastic solid and a 
counter surface, have many important applications, e.g., for rubber
wear\cite{powder}, or in adhesion and friction involving 
rubber-like materials\cite{Knaus,Crack1,Creton,adhesion1,adhesion2,Gent,Crack0}. 
Two different approaches have been applied to crack propagation in viscoelastic solids.
One focus on the stress using the cohesive-zone (or Dugdale--Barenblatt) model,
and another is based on an energy approach. The first approach was used by Knauss\cite{Knaus} and by 
Schapery\cite{Scharp1,Scharp2} and later by Hui et al\cite{Kramer1} 
and by Greenwood\cite{Green1}. Since the exact relation between the stress
$\sigma$ and the surface separation $u$ in the cohesive (process) zone, 
where the bond-breaking is assumed to occur,
is not known in general, it was
assumed that the stress is constant and equal to $\sigma_0$ for $0<u<h_0$ and $\sigma=0$ for $u>h_0$. 
The second (energy) approach was used by de Gennes\cite{Gennes} 
in a qualitative way, and by Persson and Brener\cite{Brener} in a
quantitative way. In the latter approach enters a cut-off 
radius $a_0$, which can be interpreted as the radius of curvature of the crack tip
in the adiabatic limit, and the stress $\sigma_{\rm c}$ at the crack tip 
(the stress to break the adhesive or cohesive bonds).

\vskip 0.3cm
{\bf Viscoelastic factor $G/G_0$}

The stress $\sigma_0$ in the cohesive-zone
approach is in general not the same as the stress $\sigma_c$ in the theory of Persson and Brener.
In Ref. \cite{Green1} it was assumed that the stress in the 
process zone is a constant $\sigma=\sigma_0$ for $0<z<h_0$, where 
$\sigma_0$ can be any number as long as the (adiabatic) work of adhesion is given by $G_0 = h_0 \sigma_0$.
If one use another wall-wall interaction, e.g., based on the Lennard Jones potential, 
one gets another relation between $G_0$ and $\sigma_0$. 
Since the final result depends on $\sigma_0$ this approach is in fact somewhat ill-defined
unless the exact relation between the stress and the 
separation is known (which in general is not the case) and used in the theory. 
On the contrary in the theory of Persson and Brener only well defined 
(experimental) quantities occur. Thus the adiabatic crack tip radius 
$a_0$ could in principle be measured using, e.g., 
an electron microscope, or calculated from the adiabatic work of 
adhesion $G_0$ and the stress $\sigma_{\rm c}$ to break the 
bonds via $a_0=E_0 G_0 /(2\pi \sigma_{\rm c}^2)$. In any case we expect $a_0$ to be of order
the length of the polymer chains between the cross links, i.e., typically of order $1 \ {\rm nm}$. 

\begin{figure}[tbp]
\includegraphics[width=0.25\textwidth,angle=0]{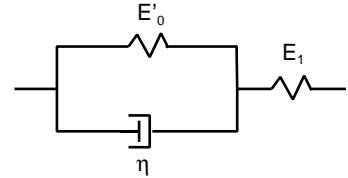}
\caption{
Three-element viscoelastic model
used in model calculation of the crack propagation energy $G(v)$.
The low frequency modulus $E(0)=E_0=E_0'E_1/(E_0'+E_1)$ 
and the high frequency modulus $E(\infty)=E_1$
and the viscosity $\eta=1/\tau$ are indicated.
}
\label{rheologypic.eps}
\end{figure}

In Ref. \cite{Brener,Crack3}, it was shown that the Persson--Brener theory gives nearly the same 
result for the viscoelastic factor $G/G_0=1+f(v)$ as the cohesive-zone model. 
Thus the numerical results for $G/G_0$ obtained by Greenwood and by Hui et al 
for the three-element viscoelastic model
(see Fig. \ref{rheologypic.eps}) is nearly the same as predicted by the Persson--Brener theory 
if one chooses $\sigma_0$ to get the best possible overlap between the two curves
(which means using $\sigma_0 \approx 3 \sigma_c$), 
see Fig. \ref{1logv.2logG.red.Greenwood.minus.0.15.Persson.plus.9.eps} 
(see Appendix A for the equations used in the calculations). 
Shifting like this is the only meaningful way 
to compare the factor $G/G_0$ between the two theories, because the velocity normalization 
factor in the cohesive-zone approach depends on $\sigma_0$ [as $v_0=E_0 G_0/(\tau \sigma_0^2$)], 
and is ill-defined (no interaction force is constant for $0<z<h_0$ and zero elsewhere).

\begin{figure}[tbp]
\includegraphics[width=0.45\textwidth,angle=0]
{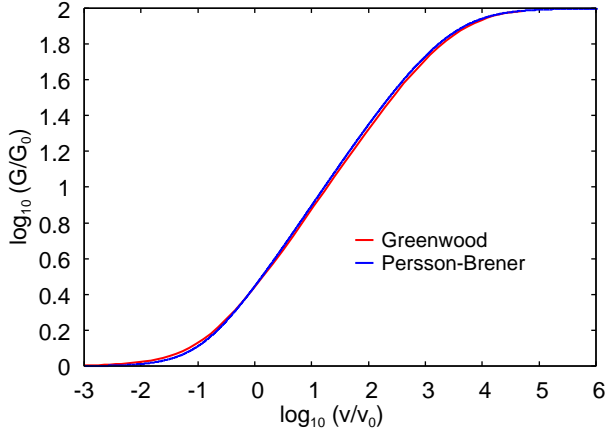}
\caption{
The crack propagation viscoelastic factor 
$G/G_0 = 1+f(v)$ as a function of the crack tip speed (log-log scale)
for the three level rheology model. The red curve is calculated by Greenwood 
using the cohesive-zone model with the 
adiabatic (infinitely slowly) work of adhesion $G_0 = h_0 \sigma_0$. The 
blue curve is from the theory of Persson and Brener assuming a 
crack (adiabatic) tip cut-off radius $a_0$. The reference velocity
$v_0 = a_0/\tau$ with $a_0=E_0 G_0 /(2\pi \sigma_{\rm c}^2)$ 
in the Persson--Brener cut-off model and 
$v_0=1.41 b/\tau$ with $b=E_0 G_0/\sigma_0^2$ in the cohesive-zone 
model.
}
\label{1logv.2logG.red.Greenwood.minus.0.15.Persson.plus.9.eps}
\end{figure}

\vskip 0.3cm
{\bf Finite size effect}

At high crack-tip speed (or at low temperatures) the main 
contribution to the viscoelastic energy dissipation
comes from a region far from the crack tip. 
This follows from dimensional arguments: the perturbing deformation frequencies
from the moving crack (velocity $v$) a distance $r$ away from 
the crack tip must be of order $v/r$. Thus close to
the crack tip the rubber will effectively be in the glassy state (elastic response) and the dominant 
contribution to the energy dissipation will 
come from regions far from the crack tip. Hence if the solid has a finite
extent, say of linear dimension $L$, at high enough 
crack tip speed the solid will, from the point of view of viscoelastic dissipation, 
effectively be in the glassy elastic state everywhere
and no viscoelastic energy dissipation will occur during the crack propagation. Thus one may be tempted
to claim that for finite solids $f \rightarrow 0$ as $v\rightarrow \infty$. This result was used
by de Gennes to argue that for the slab geometry (thickness $d_0$), 
as occur for example in pressure sensitive adhesives, for large crack tip speed
the crack propagation energy $G(v)$ will 
decrease with increasing $v$ which may result in mechanical instabilities\cite{Gennes}.
However, I will now shown in that this argument is in fact not correct and there
exist no finite size effect (see also Ref. \cite{Crack2}).

\begin{figure}[tbp]
\includegraphics[width=0.48\textwidth,angle=0]{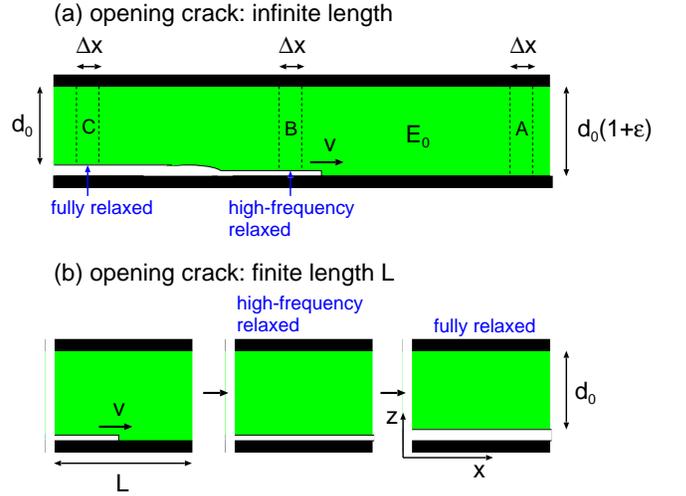}
\caption{
Crack propagation in a viscoelastic slab clamped between two rigid flat surfaces.
(a) The slab is infinite long ($L=\infty$) in the crack propagation direction ($x$-direction).
(b) The slab is of finite length $L$. In (a) the segment (of width $\Delta x$) at A is stretched with the strain $\epsilon$. At B
the strain is reduced (high frequency relaxation) by $\Delta \epsilon = \sigma_0 /E_1$ due to the abrupt decrease in the tensile stress
from $\sigma_0$ to $0$ when the crack tip pass the segment. At C, far away from the crack
tip, the strain vanish due to (slow) viscoelastic relaxation.
}
\label{newdashedopenclose1.eps}
\end{figure}

Fig. \ref{newdashedopenclose1.eps}(a) shows a fast moving opening crack in a thin viscoelastic slab 
(thickness $d_0$) under tension. The slab is assumed to be infinite long in the crack propagation direction.
The slab is elongated by $d_0 \epsilon_0$,
and we wait until a fully relaxed state is formed before inserting the crack.
Thus the elastic energy stored in the strip A of width $\Delta x$ and volume $\Delta V = w d_0 \Delta x$ (where $w$
is the width of the solid in the $y$-direction) is
$$U_0 = {1\over 2} \sigma_0 \epsilon_0 \Delta V = {\sigma_0^2 \over 2E_0} \Delta V.$$
This energy is partly used to break the interfacial bonds and partly dissipated due to the material viscoelasticity.
The crack propagation energy $G=U_0/(w \Delta x)=\sigma_0^2 d_0/(2E_0)$.

Consider now the slab $\Delta x$ as it moves from one side of the crack to the other side.
During this transition it will experience an (elongation) stress $\sigma(t)$ which for a very fast moving crack
can be considered as a step function where $\sigma = \sigma_0$ for $t<0$ and $\sigma = 0$ for $t>0$, where $t=0$ correspond to
the case where the segment $\Delta x$ is at the crack tip. The viscoelastic material will respond to this step-like change in the stress
with its high frequency modulus $E_1$ so the strain in the segment $\Delta x$ will abruptly drop by $\Delta \epsilon = \sigma_0/E_1$
as the crack pass the segment. The drop in the elastic energy  
$$\Delta U = {1\over 2} \sigma_0 \Delta \epsilon \Delta V = {\sigma_0^2 \over 2E_1} \Delta V, $$
is used to break the interfacial bonds (energy $\Delta \gamma =G_0$ per unit surface area), i.e. $G_0=\Delta U/(w \Delta x)$ or
$$G_0 = {\sigma_0^2 \over 2E_1} d_0 = {\sigma_0^2 d_0 \over 2E_0} {E_0\over E_1} = G {E_0\over E_1},$$
so that $G=G_0 E_1/E_0$. The remaining elastic energy stored in the segment $\Delta x$,
$$U= U_0-\Delta U = {1\over 2} \sigma_0^2  \left ( {1\over E_0}-{1\over E_1}\right ) \Delta V,$$
is dissipated in the slow viscoelastic relaxation occurring far away from the crack tip so that finally the material
reach its fully relaxed state (zero strain and stress). 

For a solid with a finite extent in all directions the fast relaxation process at the crack tip is the same as above so the
result $G=G_0 E_1/E_0$ still holds. However, for a fast moving crack the time it takes for the 
crack to fracture the whole interface,
$\Delta t = L/v$, is so short that, in accordance with the discussion presented earlier, 
negligible viscous energy relaxation has occurred during the crack propagation act. Thus when the crack separate the two solids
the viscoelastic solid is still in a strained state. Only after a possible long time period will it return to a strain
(and stress) free state. Thus in this case the viscoelastic energy dissipation occur in a 
process separated from the actual crack propagation [see Fig. \ref{newdashedopenclose1.eps}(b)],
and this fact was overlooked in the earlier energy based discussions of finite size effects. We conclude there exist no finite
size effects and the theory developed in Ref. \cite{Crack4,FS,Cev} is not relevant for crack propagation in viscoelastic solids. 

\begin{figure}[tbp]
\includegraphics[width=0.45\textwidth,angle=0]{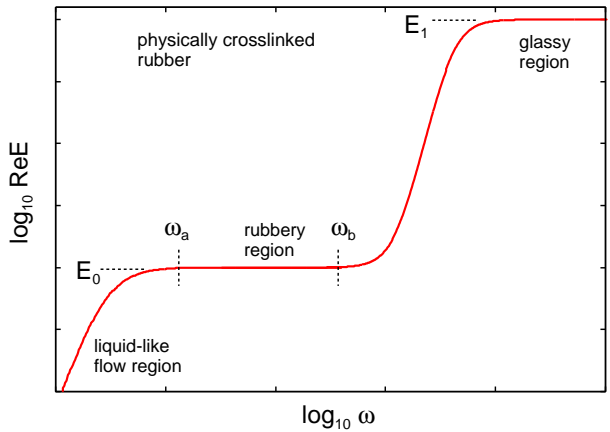}
\caption{
The real part of the viscoelastic modulus as a 
function of frequency (log-log scale) for a physically
crosslinked polymer, which behaves 
as a fluid for small frequencies, corresponding to long times (schematic).
The modulus $E_0$ in the rubbery region 
$\omega_{\rm a} < \omega < \omega_{\rm b}$ is approximately constant
and much smaller than the modulus $E_1$ in the glassy region (high frequencies).
}
\label{1logOmega.2logReE.physically.crosslinked.eps}
\end{figure}

\vskip 0.3cm
{\bf Application to ``solids'' with $E(\omega=0)=0$}

Schapery applied his viscoelastic crack propagation theory to ``solids'', with 
has a viscoelastic modulus that vanish for zero frequency (i.e., a relaxation modulus which
vanish for long times)\cite{Scharp1,Scharp2}. 
Such a ``solid'' is really a liquid with a non-Newtonian and
possible complex rheology, and in this case no rigorous crack propagation theory can be developed.
Thus, for example, the Johnson, Kendall, and Roberts (JKR) 
adhesion theory\cite{JKR} assumes that the solid deformations far from the crack
tip (here the line separating the contact area from the non-contact area), 
is characterized by the low frequency
(fully relaxed state) elastic modulus $E_0=E(\omega = 0)$, which would vanish in the present case.
We note, however, that physically crosslinked polymer materials may have a rubbery plateau 
(see Fig. \ref{1logOmega.2logReE.physically.crosslinked.eps} 
and Ref. \cite{non,xx}) for $\omega_{\rm a} < \omega < \omega_{\rm b}$, 
and behave as an elastic solid for all practical time scales 
$1/\omega_{\rm b} < t < 1/\omega_{\rm a}$, in which case
the Persson-Brener (or cohesive-zone) theory could still be applied but with $E_0$ being the modulus in the 
rubbery plateau region.  

If the adiabatic crack tip radius $a_0$ is treated as a constant as the low-frequency modulus $E_0$
is varied then the viscoelastic factor $G/G_0$ is independent of $E_0$ for low enough velocities
and hence well-defined even in the limit $E_0 = 0$ (see Appendix B). 
Nevertheless, the quantities $a_0$ and $G_0$ are not well defined in the limit $E_0=0$ since they both
refer to the adiabatic limit where the ``solid'' responds as a fluid, where no crack-like defects can occur.  

\vskip 0.3cm
{\bf Discussion}

In a recent series of papers
Ciavarella  et al \cite{C1,C2} have criticized the Persson-Brener theory 
and claimed that the Persson-Brener theory\cite{Brener} give a viscoelastic factor 
$G/G_0 = 1+f$ which differ strongly from what Ciavarella et al
denote as the exact viscoelastic factor obtained using the 
cohesive-zone model\cite{Green1,Kramer1}. We have shown above that 
both theories gives nearly the same result for $G/G_0$ if the ill-defined 
quantity $\sigma_0$ in the cohesive-zone theory is chosen appropriately!

In another paper Popov\cite{Popov} claim that viscoelasticity increases the JKR pull-off force with a factor
$E_1/E_0$ in the quasi-static limit. This result is 
incorrect: in the quasi-static
case the work of adhesion is not influenced by the viscoelasticity and only in the 
limit $v \rightarrow \infty$ is the work of adhesion increased by the factor of $E_1/E_0$. 
The result of Popov would hold if there would be no shortest length scale in the problem 
but in reality there is a short distance cut-off denoted by $a_0$ above.

\vskip 0.5cm
{\bf Acknowledgments:}
I thank J.A. Greenwood for supplying the numerical data used for the
$G(v)/G_0$ curve for the cohesive-zone model (the red curve in Fig. 1).
I thank M. M\"user for comments on the manuscript. 

\vskip 0.5cm
{\bf Appendix A}  

Here we give the equations for viscoelastic crack propagation used in the calculations
in Fig. 2.

When a strip of a viscoelastic material is exposed to an oscillating strain $\epsilon(\omega) {\rm exp}(-i\omega t)$ the
amplitude of the oscillating stress
$$\sigma(\omega )=E(\omega) \epsilon (\omega)$$
This equation define the viscoelastic modulus $E(\omega)$, which is a complex quantity, where the imaginary part is related
to energy dissipation (transfer of mechanical energy into the disordered heat motion).
In the calculation in Fig. 2 we used the three-element rheological model illustrated in Fig. \ref{rheologypic.eps}.
For this model the viscoelastic modulus
$$E= {E_0 E_1 (1-i\omega \tau) \over E_1-i\omega \tau E_0}\eqno(A1)$$
The ratio between the high frequency and low frequency modulus, $E_1/E_0$, is typically very
large, e.g., $\sim 1000$.

In the Persson-Brener theory the crack propagation energy $G(v)=G_0 a(v)/a_0$, where $a(v)$ is the
velocity dependent effective crack tip radius, and where
$a_0=E_0G_0/(2\pi \sigma_{\rm c}^2)$.
The fracture energy $G(v)$ satisfy\cite{Brener}
$${G_0\over G}= 
1- E_0  {2\over \pi} \int_0^{\omega_c} d\omega \ F(\omega) {1\over \omega} {\rm Im} {1\over E(\omega)}\eqno(A2)$$
where
$$F(\omega)=\left [1-\left ({ \omega \over \omega_c}\right )^2\right ]^{1/2}$$
where $\omega_c = 2\pi v/a$. For numerical calculations it is more convenient to reformulate (A2) into\cite{Crack1,Crack4}
$${G_0\over G} = 1- 
{E_1 {2 \over \pi}  \int_{0}^{\omega_c} d\omega {1
        \over \omega} F(\omega ) {\rm Im} {1\over E(\omega )} 
\over 1+ E_1 {2\over \pi} \int_{0}^\infty d\omega {1 \over \omega} 
{\rm Im} {1\over E( \omega )}}.\eqno(A3)$$
Since  $\omega_c$ depends on $a$ (and hence on $G$ since $G=G_0 a/a_0$) this is an implicit equation for $G(v)$.
Thus the theory gives both crack propagation energy $G(v)$ and the (velocity-dependent) radius of the crack tip,
$$a(v)= a_0 {G(v)\over G_0} = {E_0 G \over 2\pi \sigma_{\rm c}^2}. \eqno(A4)$$
Since $E(\omega )$ typically varies with $\omega$ over very many decades in frequency,
for the numerical evaluation of the integrals in (A3)
it is convenient to write (see Ref. \cite{Crack1}) $\omega = \omega_0 e^\xi$, so that if $\omega$ varies over $\sim 30$
decades, $\xi$ varies only by a factor $\sim 100$.

In the cohesive-zone model assuming 
that the stress is constant and equal to $\sigma_0$ for $0<u<h_0$ and $\sigma=0$ for $u>h_0$, and assuming the viscoelastic modulus (A1),
one get\cite{Green1}:
$${G_0\over G} = {E_0\over E_1} + {1\over 2} \left (1-{E_0\over E_1} \right ) \alpha \int_0^1 d\xi \ H(\xi ) e^{-\alpha (1-\xi)},\eqno(A5)$$
where $\alpha = l/(v\tau)$ and
$$H(\xi) = 2\xi^{1/2}-(1-\xi){\rm ln} \left ( {1+\xi^{1/2} \over 1-\xi^{1/2}} \right ),$$
and the width $l$ of the crack tip process zone
$$l = {\pi \over 4}{E_0G\over \sigma_0^2}\eqno(A6)$$
Here we have assumed plain stress; for plain strain $E$ must be replaced with $E/(1-\nu^2)$, where $\nu$ is the Poisson ratio.
Since  $l$ depends on $G$, (A5) is an implicit equation for $G(v)$.

\begin{figure}[tbp]
\includegraphics[width=0.45\textwidth,angle=0]{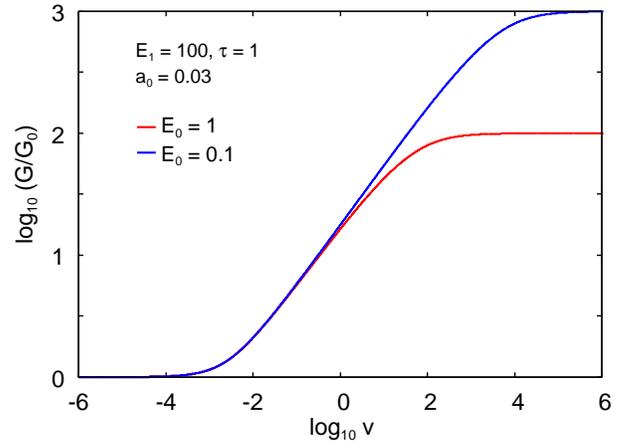}
\caption{
The viscoeladstic factor $G/G_0$ as a function of the crack tip speed (log-log scale)
for the case $E_1=100$, $\tau = 1$ and $E_0=1$ (red curve) and $E=0.1$ (blue curve).
In the calculation we have used the same adiabatic crack tip radius $a_0=0.03$. 
}
\label{1logv.2logG.E0=1.E0=0.1.E1=100.eps}
\end{figure}

\vskip 0.5cm
{\bf Appendix B}  

Using the viscoelastic modulus (A1), which we can also write as
$${1\over E} = {1\over E_1}+\left ({1\over E_0}-{1\over E_1}\right ) {1\over 1-i\omega \tau},$$
in (A2) or (A3) in Fig. \ref{1logv.2logG.E0=1.E0=0.1.E1=100.eps} 
we show the viscoelastic factor $G/G_0$ as a function of the crack tip speed
on a log-log scale. We have used $E_1=100$, $\tau = 1$ and $E_0=1$ (red curve) and $E=0.1$ (blue curve),
and the same adiabatic crack tip radius $a_0=0.03$. Note that for low crack tip
speed $G/G_0$ is independent of $E_0$ and hence well-defined also in the ``fluid'' limiting case $E_0=0$.

\end{document}